\documentclass{aa}  

\usepackage{amsmath}
\usepackage{graphicx}
\usepackage{appendix}
\usepackage{multirow}
\usepackage[varg]{txfonts}
\usepackage{natbib}
\usepackage{array}
\usepackage{nicefrac}
\bibpunct{(}{)}{;}{a}{}{,}
\usepackage[version=4]{mhchem}
\usepackage{soul}
\usepackage[dvipsnames]{xcolor}
\usepackage{ulem}

\definecolor{orange}{rgb}{1,0.5,0}

\definecolor{pink}{rgb}{0.858, 0.188, 0.478}

\definecolor{darkgreen}{rgb}{0,0.5,0}

\definecolor{lightgreen}{rgb}{0,0.5,0}

\begin{document}
\title{Broadband spectroscopy of astrophysical ice analogues}
\subtitle{II. Optical constants of CO and CO$_2$ ices
in the terahertz and infrared ranges}

\author{
    A.A.~Gavdush \inst{\ref{GPI}},
    F.~Kruczkiewicz \inst{\ref{MPE}, \ref{AMU}},\thanks{Both authors contributed equally to this work.},
    B.M.~Giuliano\inst{\ref{MPE}},
    B.~M\"uller \inst{\ref{MPE}},
    G.A.~Komandin \inst{\ref{GPI}},
    T.~Grassi \inst{\ref{MPE}},
    P.~Theulé \inst{\ref{AMU}},
    K.I.~Zaytsev \inst{\ref{GPI},\ref{BMSTU}},
    A.V.~Ivlev \inst{\ref{MPE}},
    and P.~Caselli \inst{\ref{MPE}}
}

\institute{Prokhorov General Physics Institute of
the Russian Academy of Sciences, 
119991 Moscow, Russia
\label{GPI}
\and
Max-Planck-Institut f\"ur Extraterrestrische Physik,
Gießenbachstraße 1, Garching, 85748, Germany
\label{MPE}
\and 
Aix Marseille Univ, CNRS, CNES, LAM, Marseille, France 
\label{AMU}
\and
Bauman Moscow State Technical University, 105005 Moscow, Russia
\label{BMSTU}}

\date{Received - , 2022;
Accepted - , 2022}
   
\authorrunning{A.A.~Gavdush~et~al.}

\titlerunning{Dielectric spectroscopy of
CO \& CO$_2$ ices in the THz-IR range}

\date{Received 2022; accepted  2022}

    
\abstract
    {Broadband optical constants of astrophysical ice analogues in the infrared (IR) and terahertz (THz) ranges are required for modeling the dust continuum emission and radiative transfer in dense and cold regions, where thick icy mantles are formed on the surface of dust grains.
    Such data are still missing from the literature, which can be attributed to the lack of appropriate spectroscopic systems and methods for laboratory studies.}
   {In this paper, the THz time-domain spectroscopy (TDS) and the Fourier-transform IR spectroscopy (FTIR)
   are combined to study optical constants of CO and CO$_2$ ices in the broad THz--IR spectral range.
   }
   {The measured ices are grown at cryogenic temperatures by gas deposition on a cold silicon window.
   A method to quantify the broadband THz--IR optical constants of ices is developed. It is based on the direct reconstruction of the complex refractive index of ices in the THz range from the TDS data, and the use of the Kramers-Kronig relation in the IR range for the reconstruction from the FTIR data. Uncertainties of the Kramers-Kronig relation are eliminated by merging the THz and IR spectra.
   The reconstructed THz--IR response is then analyzed using classical models of complex dielectric permittivity.}
    {The complex refractive index of CO and CO$_2$ ices deposited at the temperature of $28$~K
    is obtained in the range of $0.3$--$12.0$~THz, and fitted using the analytical Lorentz model.
    Based on the measured dielectric constants, opacities of the astrophysical dust with CO and CO$_2$ icy mantles are computed.}
    {The developed method can be used for a model-independent reconstruction of optical constants of various astrophysical ice analogs in a broad THz--IR range.
    Such data can provide important benchmarks to interpret the broadband observations from the existing and future ground-based facilities and space telescopes.
    The reported results will be useful to model sources that show a drastic molecular freeze-out, such as central regions of prestellar cores and mid-planes of protoplanetary disks, as well as CO and CO$_2$ snow lines in disks. 
    }

\maketitle

\section{Introduction}
\label{SEC:Intro}

The chemical and physical characterization of molecular clouds and protoplanetary disks, where star and planet formation is taking place, remains a challenging problem of modern astrophysics. The interplay between gas phase and icy mantles forming on the surface of dust grains can significantly affect the physical and chemical properties. In particular, catastrophic molecular freeze-out, found at the center of pre-stellar cores (e.g., \cite{Caselli_APJ_99, Caselli_ApJ_2022, Pineda__2022}), and expected in the mid-plane of protoplanetary disks (\cite{AA.338.L63.1998, FD.168.9.2014, ARAA.52.1.541.2015, Oberg_Bergin_2021}, and references therein), implies that the majority of species heavier than He reside on dust grains in these regions. Here, thick icy mantles grow around dust grains, altering dust opacities and thus the thermal balance (e.g., \cite{Keto_Caselli_2010, Hocuk2017A&A, Oka2011ApJ}), as well as profoundly affecting dust coagulation processes \citep{Chokshi1993ApJ, Dominik_Tielens_1997ApJ}. Variations in dust opacities need also to be taken into account when measuring masses from observations of the millimeter and sub-millimeter dust continuum emission.

The interpretation of observational data in the millimeter- and THz ranges by the Atacama Large Millimeter / sub-millimeter Array (ALMA) and Northern Extended Millimeter Array (NOEMA) facilities relies on the analysis of measured dust continuum emission (e.g., \cite{ARAA.57.1.79.2019, ARAA.58.1.727.2020}).
However, this analysis requires the knowledge of the dust opacity, which depends on different factors, such as the grain size distribution, their chemical composition, and the presence of ice mantles. The latter factor may critically affect characteristics of the dust opacity.

Unfortunately, available experimental data on the broadband complex dielectric permittivity (or optical  constants) of ices are quite limited. The continuum emission measurements are normally analyzed by using
model opacity values \citep{AA.261.567.1992, 1994AA.291.943O}, while most of the available experimental data provide optical constants of ices in the visible and mid-IR ranges
\citep{AJ.86.713.1993, AA.328.649.1997, JOSAA.15.12.3076.1998, AA.435.2.587.2005, AA.445.3.959.2006, PCCP.8.2.279.2006, JGRA.113.D14.D14220.2008, AJ.701.2.1347.2009}.

Far-IR spectroscopy of pure molecular ices and their mixtures has so far been performed with the aim of measuring the IR-active lattice vibrations of amorphous and crystalline phases of astrophysical ice analogues \citep{JCP.45.11.4359.1966, JCP.46.10.3991.1967, 1992ApJ...401..353M, AA.103.45.1994}, without deriving their optical constants. The band strengths have been investigated by using the Fourier-transform IR (FTIR) spectroscopy \citep{AA.565.A108.2014, AA.592.A81.2016}. Also, THz optical constants of astrophysical ice analogs have been explored in experiments, demonstrating the ability of spectroscopy in this frequency range to provide important information on the lattice structure, large-scale structural changes, and thermal history of ices \citep{PCCP.16.8.3442.2014, FD.168.461.2014, PCCP.18.30.20199.2016, FASS.8.757619.2021}.

Recently, \cite{AA.629.A112.2019} developed a new approach for quantitative model-independent measurements of the THz complex dielectric permittivity of ices grown at cryogenic conditions. They used the THz time-domain spectroscopy (TDS) in the transmission mode, which provides detection of {\it both} the amplitude and phase of the THz signal, obtained in a single rapid measurement.
In contrast to FTIR spectroscopy, TDS enables reconstruction of the complex dielectric response of a given ice sample directly from the measured data, without using the Kramers-Kronig relation or employing additional assumptions \citep{AA.629.A112.2019, OME.10.9.2100.2020, OE.30.6.9208.2022}.
The developed method was applied by \cite{AA.629.A112.2019} to quantify the dielectric response of CO ice in the $0.3-2.0$~THz range. The obtained results shows good agreement with available data on the refractive index of CO ice in the far-IR spectral range.

While independent THz and IR measurements of astrophysical ice analogs are well known from the literature, their broadband (THz--IR) characterization still remains challenging. In this paper, we extend the spectral range of quantification of the complex dielectric permittivity, by merging TDS and FTIR data.
For this purpose, we combine the TDS and the FTIR spectrometers to the same cryogenic setup, which used to grow the ice analogs under identical conditions (see Section~\ref{SEC:ExpMeth}). Then, we apply a newly developed algorithm, which allows us to reconstruct the complex dielectric permittivity of an ice sample in the range of $0.3-12.0$~THz, thus considerably extending the frequency range analyzed in \cite{AA.629.A112.2019}. This algorithm ensures efficient elimination of well-known uncertainties associated with the Kramers-Kronig relation, as discussed in Section~\ref{SEC:Res}. We use the developed method to study the optical constants of CO and CO$_{2}$ ices at $11$~K. The relevance of the obtained data to the astrophysical applications is highlighted by calculating opacity of the astrophysical dust covered with thick CO and CO$_2$ ice mantles. The results are presented in Section~\ref{SEC:Results}. Our findings can now be applied for interpretation of astronomical observations in the THz--IR range.

\section{Experimental procedure}
\label{SEC:ExpMeth}

The experimental data were acquired at the CASICE laboratory developed at the Center for Astrochemical Studies located at the Max Planck Institute for Extraterrestrial Physics in Garching (Germany). In this work we combine two different instruments of the laboratory, a TDS spectrometer and a FTIR spectrometer, to obtain broadband optical constants of CO and CO$_2$ ices.

The experimental procedure, including details of the TDS instrument, has been discussed in our first publication by \cite{AA.629.A112.2019}, while details of the FTIR instrument can be found in \cite{Mueller+2018}. 
In this section, we briefly summarize the main points of the experimental setup and its operation, with an emphasis on the ice sample preparation for the TDS and FTIR measurements.

\subsection{The experimental chamber}
\label{SEC:Cryo}

The design of the TDS and FTIR instruments allows us to use the same vacuum chamber for growing the ice samples, and therefore we can switch between the two beams. As a result, the ice samples used for TDS and FTIR measurements have reproducible properties (for given deposition conditions), which makes it possible to merge THz and IR data and thus to obtain broadband optical constants of ices.

\begin{figure*}
   \centering
   \includegraphics[width=1.9\columnwidth]{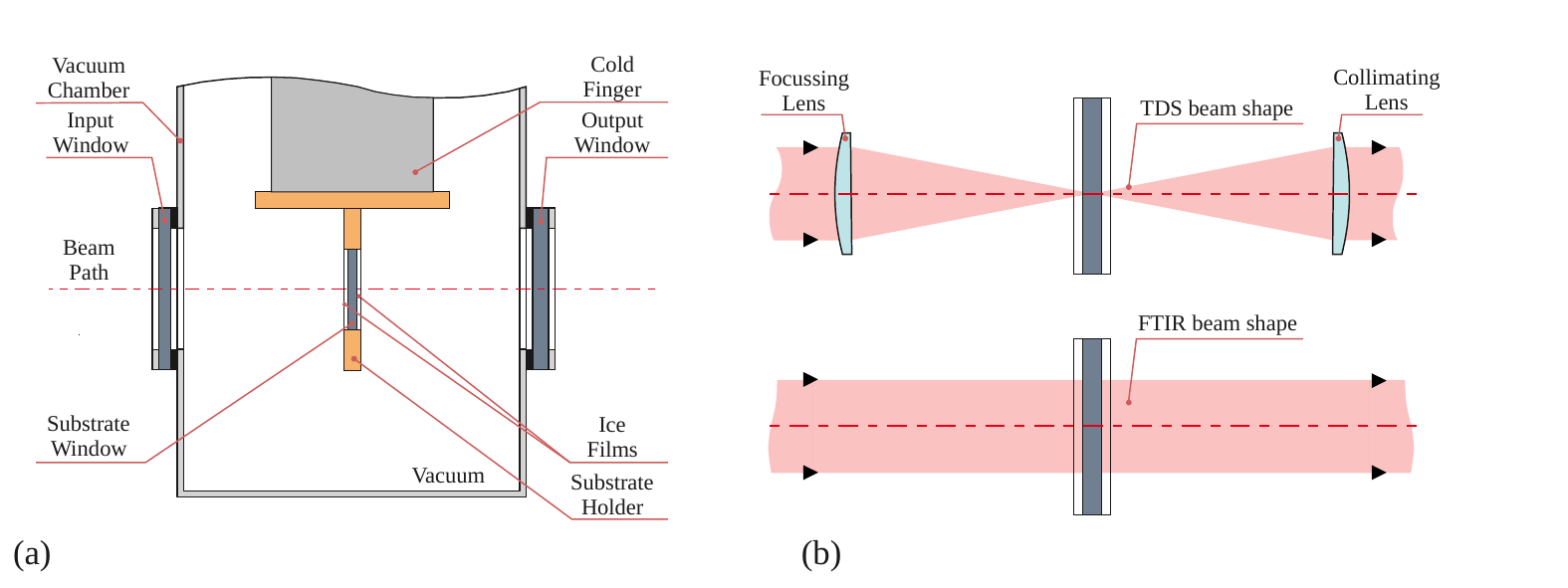}
   \caption{Sketch of the experimental setup,
   showing the optical system, the vacuum chamber and cryocooler arrangement.
   (a)~Bottom part of the vacuum chamber, showing the cold finger with the sample holder 
   and ice films on it.
   (b)~Optical paths of the TDS and FTIR beams through the ice samples. Panel~(a) is adapted from
   \citep{AA.629.A112.2019}.}
   \label{FIG:Setup}
\end{figure*}

The vacuum chamber is mounted on a motor-controlled translation stage, which ensures tuning of the cryostat position with respect to the THz beam, and also allows us to move the chamber and perform measurements with the FTIR spectrometer. The chamber has a diameter of $15$~cm, and can be hosted in the sample compartment of both spectrometers. It is equipped with a high-power closed-cycle cryocooler (Advanced Research Systems). A sketch of the vacuum chamber and the optical arrangement of the TDS and FTIR beams are shown in Fig.~\ref{FIG:Setup}.

The minimum measured temperature that can be reached at the sample holder in normal operation mode is $5$~K. For this set of experiments, a special configuration has been chosen in order to provide homogeneous deposition of ice on the cold substrate. For this purpose, the radiation shield was removed from the sample holder, and therefore the minimum temperature achievable was $11$~K. The pumping station, composed of a turbo-molecular pump combined with a backing rotary pump, sets a base pressure of about $10^{-7}$~mbar 
when cold. The degree of water contamination in our setup was estimated from previous dedicated test experiments, showing the formation of water ice on top of the cold substrate at a rate between 20 to 60 monolayers per hour.

The optical windows and the substrate chosen for the measurements in the THz--IR range are made of high-resistivity float-zone silicon (HRFZ-Si). This material has a high refractive index of $n_{\rm Si} \approx 3.4$, 
with negligible dispersion and good transparency in the desired frequency range. The silicon substrate is placed in the middle of the vacuum chamber and is mechanically and thermally connected to the cryostat. 

\subsection{TDS and FTIR spectrometers}
\label{SEC:Specs}

The experimental setup includes a TDS system (BATOP TDS--$1008$), with a customized sample compartment to allocate the cryocooler during the THz measurements.
The TDS beam is generated by two photo-conductive antennas, which constitute the emitter and detector of the THz pulse, triggered by a femtosecond laser (TOPTICA). The TDS features a broadband spectrum spanning the range of up to $0.05$--$3.5$~THz, the spectral maximum at $\approx 1.0$~THz, and the spectral resolution down to $\approx 0.03$~THz.
The TDS housing is kept under purging with cold nitrogen gas during the entire experiment, to mitigate the absorption features due to the presence of atmospheric water in the THz beam path.

Transmission IR spectra of the ices are recorded using a high-resolution Bruker IFS $120/5$HR FTIR spectrometer. We choose a resolution of $1$~cm$^{-1}$ and $64$~scans taken per spectra. To select the wavelength range in the FIR and mid-IR, we work with a Mylar Multilayer beam splitter, a FIR-Hg source and a FIR-DTGS detector.
The sample compartment of our FTIR system is kept under vacuum, with a customized flange to accommodate the cryocooler vacuum chamber during the measurements.

\subsection{The experimental protocol}
\label{SEC:IcePrep}

To prepare ices, we used a standard procedure, keeping the same experimental conditions for both TDS and FTIR measurements. CO or CO$_2$ gas is introduced into the cryocooler vacuum chamber through a $6$-mm-diameter stainless steel pipe, with a given gas flux controlled by a metering valve. The gas expands inside the chamber and condensates onto the cold substrate, forming ice films on both sides of the substrate. 

In Sec.~\ref{SEC:Res} we demonstrate that reliable reconstruction of the broadband THz-IR optical constants
requires ice layer thicknesses of (at least) a few tenths of mm. 
To grow such a thick ice within a reasonable experimental time, we choose fast deposition conditions, in which a considerable amount of gas is allowed in the chamber where the pressure was kept at $\approx10^{-3}$~mbar during the deposition. For both instruments, we collected a series of ice spectra after each step of $3$-min-long deposition.

In order to grow ice samples with good optical properties at the desired thickness \citep[see][for details]{AA.629.A112.2019}, the gas inlet has been custom-designed. The inlet pipe is kept at a distance of $\approx 7$~cm from the substrate. With this configuration, we expect to deposit ice layers of high uniformity on both sides of the substrate.

The temperature of cold substrate before the deposition is kept at $11$~K. However, during deposition the temperature increases due to gas condensation onto the substrate, leading to the surface heating rate which is too high to be efficiently compensated by the cooling system.
The maximum temperature in the end of each deposition step is $\approx 28$~K. The system is allowed to thermally equilibrate between the deposition steps, until the substrate temperature returns to $11$~K. 

Before starting the ice deposition, the TDS and FTIR transmission spectra of bare substrate were collected. These {\it reference} spectra have then been used for deriving the optical constants of ice samples, as discussed below.

\section{TDS and FTIR data processing}
\label{SEC:Res}

At the first step, we apply apodization procedure (window filtering) to
all measured TDS waveforms and FTIR interferograms. Tukey window \citep{Tukey.1986} is used for the TDS data, 
as described in \cite{AA.629.A112.2019}. For the FTIR data, the $4^\mathrm{th}$-order Blackman-Harris window is employed \citep{ProcIEEE.66.1.51.1978}, in order to filter out the side maxima of the interferogram
while keeping a spectral resolution of $\approx 1$~cm$^{-1}$.

\begin{figure*}
    \centering
    \includegraphics[width=1.9\columnwidth]{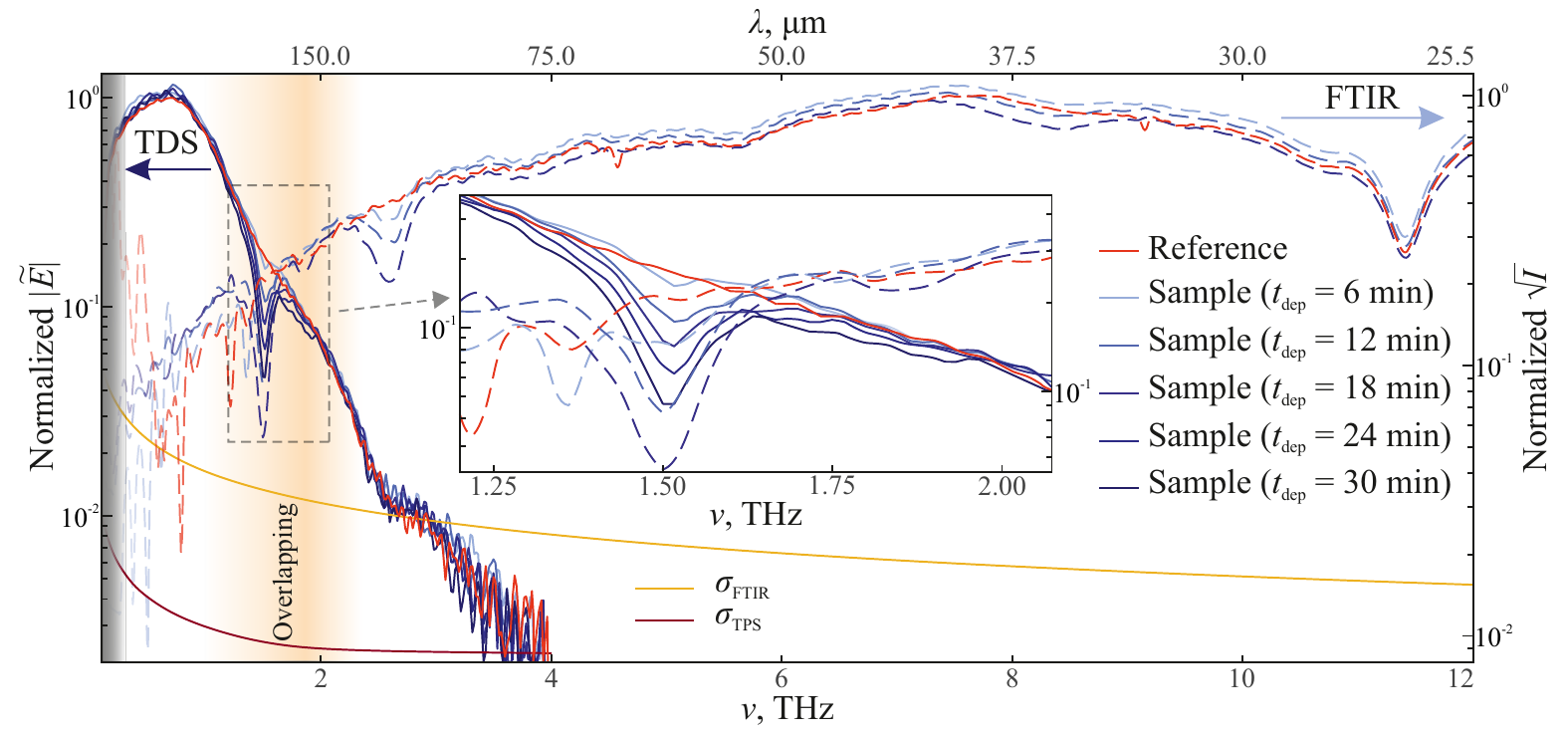}
    \caption{Reference spectrum and sample spectra (by field) of CO ice films, measured by the TDS and FTIR systems at different deposition steps (and normalized to the maximum of the respective reference spectrum for convenience).
    The grey shading toward the lower frequencies shows the spectral range where diffraction distortions of the THz beam at the sample aperture are expected to become significant, as discussed by \cite{AA.629.A112.2019}.
    The orange shading indicates qualitatively the spectral range of the TDS and FTIR data overlap used for their merging (see Sec.~\ref{SEC:Merg}). 
    Sensitivity of the TDS and FTIR measurements is characterized by the standard deviation of the respective instrumental noise, $\sigma_\mathrm{TDS}$ and $\sigma_\mathrm{FTIR}$ (see text for details).}
    \label{FIG:SampleRefSp}
\end{figure*}

Different principles underlying the TDS and FTIR spectroscopy imply different quality of physically relevant information contained in their signals.
TDS detects the time-dependent amplitude of electric field $E \left( t \right)$, from which the frequency-domain {\it complex} amplitude
$\widetilde{E} \left( \nu \right)$ is calculated via the Fourier transform, thus providing both the amplitude and phase information.
As shown in \cite{AA.629.A112.2019}, this method enables direct reconstruction of the complex dielectric permittivity of ice samples in the THz range as well as an accurate assessment of their thickness.
On the other hand, FTIR spectroscopy generates time/spatial-domain interferograms $I \left( t \right)$, related (via the Fourier transform) to the power spectrum $I \left( \nu \right) \propto \left| \widetilde{E} \left( \nu \right) \right|^{2}$ \citep{GriffithsFTIR_Book1986}.
The latter contains only the amplitude information, and thus the complex dielectric permittivity cannot be directly reconstructed.

Characteristic reference and sample spectra, measured with the TDS and FTIR systems for CO ices of different thicknesses (proportional to the deposition time), are depicted in Fig.~\ref{FIG:SampleRefSp}. We notice a broad overlap of the TDS and FTIR data at frequencies near $2.0$~THz. The sample spectra show significant evolution with the deposition time, revealing absorption features at different frequencies. 
We also plot the sensitivity curves for the TDS and FTIR measurements, represented by the respective frequency-dependent standard deviations $\sigma\left( \nu \right)$ which were estimated from the variability of TDS and FTIR reference spectra,
\begin{equation}
    \begin{split}
        \sigma_\mathrm{TDS} \left( \nu \right)
        &= \sqrt{\frac{1}{N-1} \sum_{j=1}^{N}  
        \left( \left|\widetilde{E}_{\mathrm{r},j} \left( \nu \right)  \right|
        - \langle\left|\widetilde{E}_\mathrm{r} \left( \nu \right)  \right|\rangle \right)^2 }\:,\\
        \sigma_\mathrm{FTIR} \left( \nu \right)
        &=
        \sqrt{\frac{1}{N-1} \sum_{j=1}^{N}  
        \left( \sqrt{I_{\mathrm{r},j} \left( \nu \right)}
        - \langle\sqrt{ I_\mathrm{r} \left( \nu \right) }\:\rangle \right)^2 }\:,
    \end{split}
    \label{EQ:SigmaTPS}
\end{equation}
where
\begin{equation}
        \langle\left|\widetilde{E}_\mathrm{r} \left( \nu \right)  \right|\rangle = \frac{1}{N} \sum_{j=1}^{N} \left|\widetilde{E}_{\mathrm{r},j} \left( \nu \right)  \right|,
        \quad
        \langle\sqrt{I_\mathrm{r} \left( \nu \right) }\:\rangle
        = \frac{1}{N} \sum_{j=1}^{N} \sqrt{ I_{\mathrm{r},j} \left( \nu \right) }\:,
\end{equation}
are the mean reference spectra of TDS and FTIR systems, respectively, calculated from $N$ independent reference signals (measured for different experiments). 
To facilitate further analysis, the sensitivity curves are fitted by the power-law dependence $\sigma \left( \nu \right) = A \nu^{a}$, with $A=0.002\pm 10^{-3}$, $a=0.60\pm 10^{-2}$ for the TDS system, and $A=0.04\pm 10^{-3}$, $a=0.39\pm 10^{-2}$ for the FTIR system.

The collected reference and sample signals allow us to compute the  transmission coefficients of ices. TDS data yield the complex transmission coefficient (by field),
\begin{equation}
    \widetilde{T}_\mathrm{TDS}
    \left( \nu \right) =
    \frac{ \widetilde{E}_\mathrm{s}
    \left( \nu \right) }
    { \widetilde{E}_\mathrm{r}
    \left( \nu \right) }\:,
    \label{EQ:TPSTransmission}
\end{equation}
while FTIR data provide only its amplitude,
\begin{equation}
    \left| \widetilde{T}_\mathrm{FTIR}
    \left( \nu \right) \right| =
    \sqrt {\frac{ I_\mathrm{s}
    \left( \nu \right) }
    { I_\mathrm{r}
    \left( \nu \right) } }\:.
    \label{EQ:FTIRTransmission}
\end{equation}
At the next step, we have to merge these transmission spectra, i.e., (i) the FTIR phase must be reconstructed, and (ii) the TDS and FTIR amplitudes and phases must be matched.  

\subsection{Merging TDS and FTIR data}
\label{SEC:Merg}

\begin{figure}[!b]
    \centering
    \includegraphics[width=1.0\columnwidth]{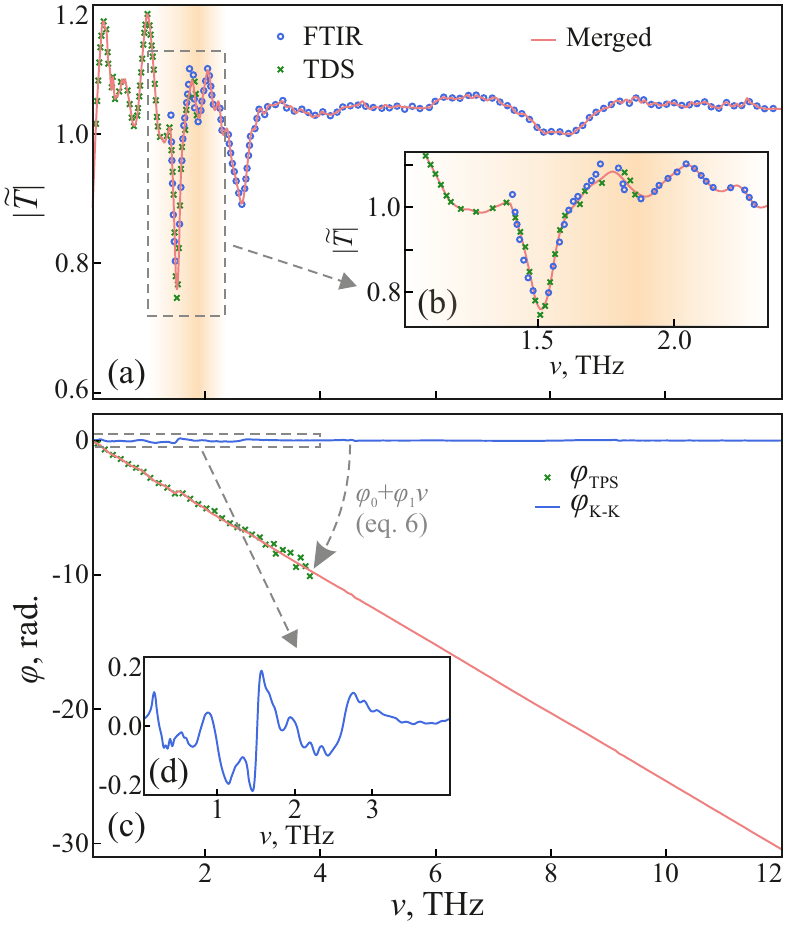}
    \caption{Merging procedure of the TDS and FTIR data, illustrated for CO ice after $12$-min deposition.
    (a)~Amplitude of the transmission coefficient
    $\left| \widetilde{T} \left( \nu \right) \right|$
    from TDS measurements (green crosses) and FTIR measurements (blue circles)
    as well as the resulting merged curve
    (red solid line, see Sec.~\ref{SEC:Merg} for details).
    (b)~Zoom-in on the overlapping range of the TDS and FTIR data.
    (c)~Phase $\phi\left( \nu \right)$ of the complex transmission coefficient, where the TDS data, the Kramers-Kronig phase (blue solid line) with the linear correction (see Eq.~6), and the resulting broadband phase (red solid line) are plotted.
    (d)~Low-frequency behavior of the Kramers-Kronig phase.}
    \label{FIG:MergedTransmission}
\end{figure}

Merging the THz and IR transmission amplitude
and reconstruction of the broadband (THz--IR) phase
are carried out independently.
From Fig.~\ref{FIG:MergedTransmission}~(a)
we notice that
the TDS  (green crosses) and FTIR (blue circles) transmission amplitudes
occur to be almost identical in the spectral range where the data overlap. 
The broadband transmission amplitude
$\left| \widetilde{T} \left( \nu \right) \right|$
is calculated using a weighted superposition
of the TDS and FTIR data in the overlapping range,
based on  frequency-dependent signal-to-noise ratios for both systems.
We tried different methods to merge the data, 
based on different ways to account 
for contributions of the THz and IR spectra in the resultant curve:
in particular, a linear weighting of data 
in a varying range of frequencies (within the overlapping range) was tested.
All these methods lead to practically indistinguishable results.
Calculation of the merged transmission amplitude
$\left| T \left( \nu \right) \right|$
is illustrated in
Figs.~\ref{FIG:MergedTransmission}~(a) and~(b)
for a CO ice sample.

In principle, the missing IR phase can be reconstructed from 
the amplitude of FTIR transmission. Consider the following logarithmic representation
of a complex transmission coefficient:
\begin{equation}
    \ln \left( \widetilde{T} \left( \nu \right) \right)
    = \ln \left| \widetilde{T} \left( \nu \right) \right|
    + i \phi \left( \nu \right).
    \label{EQ:LogarithmicTransmission}
\end{equation}
As a response function of a physical system, the
real part $\ln \left| T \left( \nu \right) \right|$
and the imaginary part $\phi \left( \nu \right)$ 
of this logarithmic transmission
are connected via the Kramers-Kronig relations 
\citep[the Gilbert transform, see][]{PR.161.1.143.1967, AIPAdv.2.3.032144.2012}.
However, the resulting phase function $\phi_\mathrm{K-K} \left( \nu \right)$
is determined with some uncertainty 
\citep{AIPAdv.2.3.032144.2012}.
This uncertainty can be corrected by writing the relation
between the desired
$\phi_\mathrm{FTIR} \left( \nu \right)$
and the calculated
$\phi_\mathrm{K-K} \left( \nu \right)$ phases in the following form:
\begin{equation}
    \begin{split}
    \phi_\mathrm{FTIR} \left( \nu \right)
    = \phi_\mathrm{K-K} \left( \nu \right) +\phi_0 + \phi_1 \nu,
    \end{split}
    \label{EQ:PhaseUncertinity}
\end{equation}

where $\phi_{0}$ and $\phi_{1}$ are constants. Notice that
$\phi_\mathrm{K-K} \left( \nu \right)$
reproduces the shape of a desired phase,
that originates from
both the material dispersion
and the interference effects,
while the correction term $\phi_0 + \phi_1 \nu$
requires additional explanation.
When the Kramers-Kronig relation
is used to retrieve the real part $n$ of
the complex refractive index $\widetilde{n}$
(based on the measured imaginary part),
it leads to uncertainty which can be presented as $n + C$.
By applying it to retrieve the phase of
the complex transmission coefficient $\widetilde{T}$, this yields
$\phi_\mathrm{K-K} \approx - i 2 \pi \nu \left( n + C - 1 \right) l / c_\mathrm{0}$, where $l$ is the total ice thickness and
$c_\mathrm{0} \approx 3 \times 10^{8}$~m/s 

is the speed of light in free space.
This results in the correction term $\phi_{1}\nu$
in Eq.~\eqref{EQ:PhaseUncertinity}.
Furthermore, since the actual phase at low frequencies, inaccessible for measurements, is unknown, a constant phase shift $\phi_0$ must be added as well. 
Figures~\ref{FIG:MergedTransmission}~(c) and~(d)
illustrate the procedure of calculating the broadband phase $\phi \left( \nu \right)$ for a CO ice sample. The constants $\phi_{0,1}$ are computed from least-squares method for Eq.~\eqref{EQ:PhaseUncertinity} and a discrete set of measured TDS phases in the overlapping range. The broadband phase is then obtained by merging
$\phi_\mathrm{FTIR} \left( \nu \right)$ and $\phi_\mathrm{TDS} \left( \nu \right)$,
in a similar manner as described above for the amplitudes. 

\subsection{Reconstruction of
the broadband dielectric response}

The broadband transmission amplitude
$\left| T \left( \nu \right) \right|$
and phase
$\phi \left( \nu \right)$
are then applied
for the reconstruction of
thicknesses and complex dielectric permittivity of
ice layers
in the frequency range of $0.3$--$12.0$~THz,
as described by
\cite{AA.629.A112.2019}.
The complex dielectric permittivity of ices
$\widetilde{\varepsilon} \left( \nu \right)
= \varepsilon' \left( \nu \right)
- i \varepsilon'' \left( \nu \right)$,
with its real $\varepsilon'$ and imaginary $\varepsilon''$ parts,
is estimated via the minimization of an error functional,
that quantifies a discrepancy between
the measured complex transmission coefficient
$\widetilde{T} \left( \nu \right)$
and that from the theoretical model.
As discussed in detail by \cite{AA.629.A112.2019},
our theoretical model of the interaction of electromagnetic waves with ice samples is generic, 
and therefore is conceptually applicable both for the THz and IR frequencies. 
Certain parameters of the model must be tuned for the IR range, including
the apodization filter size and the number of considered satellite pulses.
We remind that the satellite pulses, pronounced in the TDS data,
allow us to accurately determine
the thicknesses of ice layers
\citep[see Fig.~4 in][]{AA.629.A112.2019}. 
However, the satellites occur to be suppressed in FTIR data,
which may be attributed to an enhanced surface scattering
(as compared to the THz wavelengths)
as well as to the applied
Blackman-Harris FTIR interferogram apodization,
reducing the signal at larger delays
(as compared to the Tukey apodization of the TDS waveforms).

In Sec.~\ref{SEC:Results}, we express the complex dielectric response of ices
in terms of the complex refractive index (optical constants):
$\widetilde{n} \left( \nu \right)
    = n \left( \nu \right)
    - i c_\mathrm{0} \alpha \left( \nu \right) / \left( 2 \pi \nu \right)
    \equiv \sqrt{ \widetilde{ \varepsilon } \left( \nu \right) }\:,$
where $n$ and $\alpha$ are the refractive index
and the absorption coefficient (by field), respectively.

\subsection{Modeling of the broadband dielectric response}
\label{SEC:DielModel}

The resonance dipole excitations,
underlying the broadband dielectric response of ices,
are modeled by applying methods of dielectric spectroscopy.
Instead of the Gaussian bands, commonly used to fit
the absorption peaks of astrophysical ices
\citep[e.g.,][]{ARAA.52.1.541.2015},
we employ a physically motivated superposition of $N_\mathrm{L}$ Lorentz kernels,
\begin{equation}
    \widetilde{\varepsilon}\left( \nu \right)
    = \varepsilon_\infty
    + \sum_{j=1}^{N_\mathrm{L}} \frac{ \Delta\varepsilon_j\nu_{\mathrm{L},j}^{2} }
    { \nu^{2}_{\mathrm{L},j} - \nu^{2}  + i \nu \gamma_{\mathrm{L},j} }\:,
    \label{EQ:LorentzModel}
\end{equation}
where $\Delta\varepsilon_j$ is an amplitude,
$\nu_{\mathrm{L},j}$ is a resonance frequency,
and $\gamma_{\mathrm{L},j}$ is a damping constant
of the $j^\mathrm{th}$ Lorentz term,
whereas $\varepsilon_\infty$ is the (real) dielectric permittivity
at high frequencies
(well above the analyzed spectral range).
The magnitude $\Delta \varepsilon_j$ of each Lorentz oscillator
regulates its contribution
to the resultant complex dielectric function,
while the resonance frequency $\nu_{\mathrm{L},j}$
and damping constant $\gamma_{\mathrm{L},j}$
define its spectral position and bandwidth.

The reason for choosing the multi-peak Lorentz model
of Eq.~\eqref{EQ:LorentzModel} is twofold:
(i) it simultaneously describes the real and imaginary parts
of the complex dielectric permittivity $\widetilde{\varepsilon}$
with a minimum number of physical parameters, 
and (ii) it obeys both the sum rule
\citep{PR.161.1.143.1967,OE.30.6.9208.2022}
and the Kramers-Kronig relations
\citep{PR.161.1.143.1967, AIPAdv.2.3.032144.2012}.
The latter is a necessary requirement for self-consistent models of the permittivity function; we note that models based on
Gaussian fitting of the absorption peaks
fail to address this requirement.

To fit the experimental dielectric curves
with Eq.~\eqref{EQ:LorentzModel}, we set
the number $N_\mathrm{L}$ of the resonant lines detected 
in the ice absorption spectrum.
Then, the peak positions of these lines
are estimated, as a first approximation
for the resonance frequencies
$\nu_{\mathrm{L},j}$.
Finally, the measured dielectric curves
are fitted with the model,
using the nonlinear solver
based on an interior point algorithm
\citep{SIAMJO.9.4.877.1999}.

\subsection{Calculation of the detectable absorption values}

Along with strong resonant spectral features, astrophysical ice analogs may exhibit some relatively small frequency-dependent absorption of $\alpha \sim 1$~cm$^{-1}$ between the peaks
\citep{AA.629.A112.2019}, which appears to be close to the sensitivity limit of
our setup. To quantify detectable values of $\alpha\left( \nu \right)$ for the TDS and FTIR systems,
we first introduce the respective $3\sigma$ detection limits $\delta T\left( \nu \right)$ for the sample transmission drop:
\begin{equation}
    \delta T_\mathrm{TDS} \left( \nu \right) 
    = \frac{ 3\sigma_\mathrm{TDS} \left( \nu \right) }{ \left| E_\mathbf{r} \left( \nu \right) \right| }\:,
    \qquad
    \delta T_\mathrm{FTIR} \left( \nu \right) 
    = \frac{ 3\sigma_\mathrm{FTIR} \left( \nu \right) }{\sqrt{I_\mathbf{r} \left( \nu \right)} }\:.
    \label{EQ:DetectableTransmissionDrop}
\end{equation}
By using the Bouguer‐Beer‐Lambert law for a rough approximation of the detection limit,
$\delta T \left( \nu \right)\approx 1-
    \exp \left(-\delta \alpha \left( \nu \right) l \right),$
and applying the Taylor expansion,
we obtain the corresponding $3\sigma$ detection limits $\delta\alpha\left( \nu \right)$
for the absorption coefficient:
\begin{equation}
    \delta \alpha_\mathrm{TDS} \left( \nu \right)
    \approx \frac{ 3\sigma_\mathrm{TDS} \left( \nu \right) }{ l \left| E_\mathbf{r} \left( \nu \right) \right| }\:,
    \qquad
    \delta \alpha_\mathrm{FTIR} \left( \nu \right)
    \approx \frac{ 3\sigma_\mathrm{FTIR} \left( \nu \right) }{ l \sqrt{I_\mathbf{r} \left( \nu \right)}}\:.
    \label{EQ:DetectableAbsorption}
\end{equation}

\section{Results}
\label{SEC:Results}

\subsection{Broadband dielectric response of
CO and CO$_{2}$ ices}

\begin{figure}[!t]
    \centering
    \includegraphics[width=1.0\columnwidth]{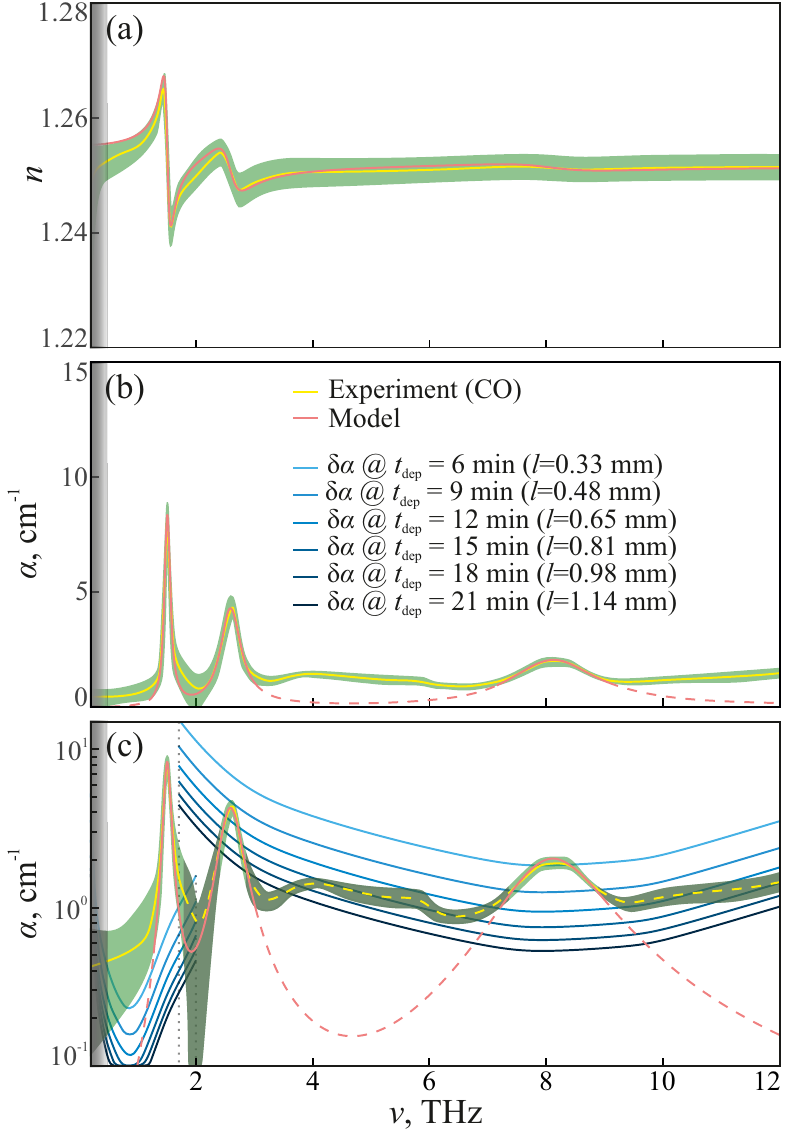}
    \caption{Broadband THz--IR optical constants of the CO ice versus the frequency $\nu$, deduced from measurements of ice films of different thicknesses. 
    The yellow solid lines show the mean values and the green shaded zones are the $\pm 1.5\sigma$ ($87$\%) confidence intervals of the measurements, while the red solid lines represent the dielectric permittivity model (see Eq.~\eqref{EQ:LorentzModel} and Table~\ref{TAB}).
    (a)~Refractive index $n$; (b,~c)~absorption coefficient $\alpha$ (by field), plotted in the linear and logarithmic scales, respectively.
    In~(c), the blue-shaded solid lines show $3\sigma$ detection limits of absorption, $\delta\alpha$, 
    estimated from Eq.~\eqref{EQ:DetectableAbsorption} for different ice thicknesses $l$. The uncertainty of $l$ ($\sim0.01$~mm, see text for details) is practically independent of the thickness.}
    \label{FIG:COResponse}
\end{figure}

\begin{figure}[!t]
    \centering
    \includegraphics[width=1.0\columnwidth]{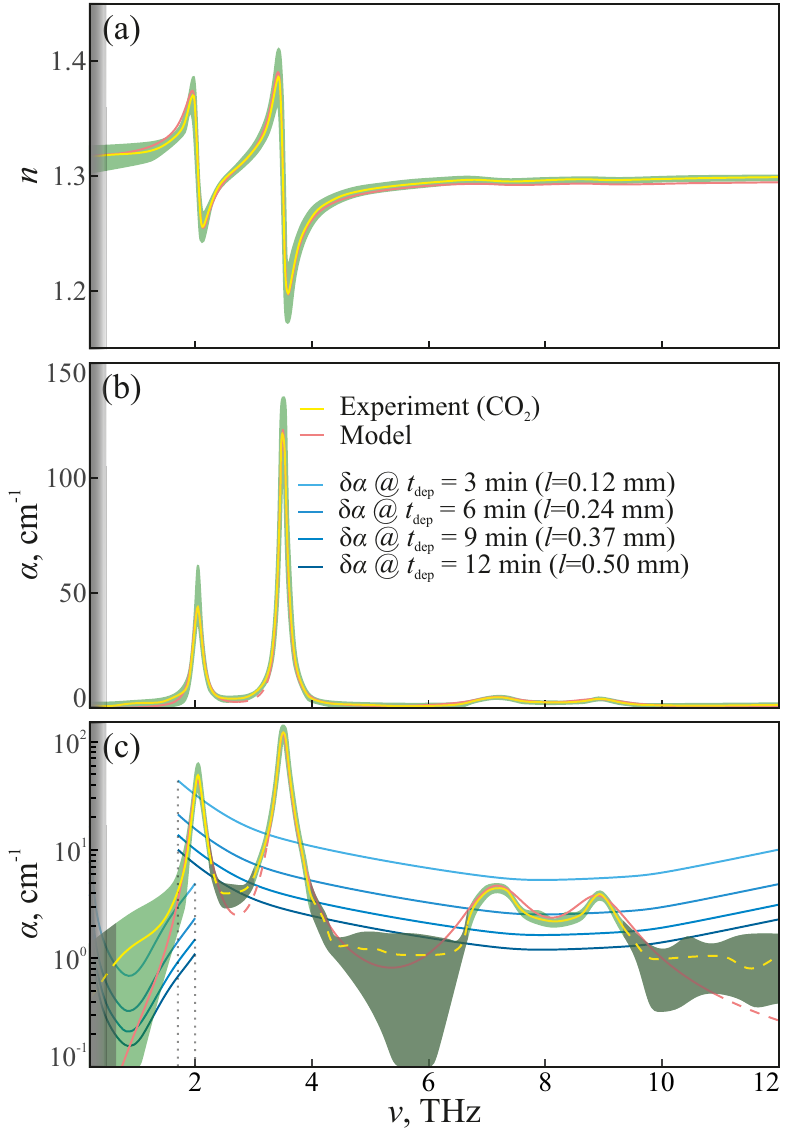}
    \caption{Same as Fig.~\ref{FIG:CO2Response} for the CO$_{2}$ ice.}
    \label{FIG:CO2Response}
\end{figure}

The described approach was applied to study
the optical constants of CO and CO$_2$ ices,
deposited at a temperature of $28$~K.
The observed results are shown in
Figs.~\ref{FIG:COResponse} and~\ref{FIG:CO2Response}. By analyzing the dielectric response of ices
at different deposition steps
and, thus, different thicknesses of ice layers,
the average optical constants
and their standard deviations
($\pm 1.5 \sigma$ or $87\%$) were estimated,
as shown in the figures by the yellow solid lines
and green shaded areas, respectively.
The obtained THz-IR response
was modeled by Eq.~\eqref{EQ:LorentzModel},
with the resultant parameters
summarized in Table~\ref{TAB}.
The blue-shaded lines
in Figs.~\ref{FIG:COResponse}~(c)
and~\ref{FIG:CO2Response}~(c)
show the $3\sigma$ detection limit $\delta \alpha$,
calculated for different ice thicknesses $l$
from Eq.~\eqref{EQ:DetectableAbsorption}.
The accuracy of \textit{in situ} ice thickness determination can be roughly estimated \citep[see, e.g.,][]{OL.22.12.904.1997, OE.52.6.068203.2013} as $\sim10\%$ of the shortest wavelengths contributing to the detectable TDS spectrum ($\sim100~\mu$m, see Fig~\ref{FIG:SampleRefSp}), which yields the uncertainty of $\sim0.01$~mm.

\begin{table}[!t]
    \caption{Parameters of the broadband dielectric permittivity model for CO and CO$_2$ ices, Eq.~\eqref{EQ:LorentzModel}, with $\pm1\sigma$ confidence intervals.}
    \centering 
        \begin{tabular}{ >{\centering\arraybackslash}m{2.0cm}
                        |>{\centering\arraybackslash}m{2.5cm}
                        |>{\centering\arraybackslash}m{2.5cm}}
            \rule[-1ex]{0pt}{3.5ex}    {Parameter}
                                     & {CO}
                                     & {CO$_2$}\\
            \hline \hline
            \rule[-1ex]{0pt}{3.5ex}    {$\varepsilon_\infty$}
                                     & {$1.56 \pm 0.01$}
                                     & {$1.67 \pm 0.02$}\\
            \rule[-1ex]{0pt}{3.5ex}    {$\Delta\varepsilon_\mathrm{1}$}
                                     & {$(55.0 \pm 0.1) \times 10^{-4}$}
                                     & {$(273 \pm 8) \times 10^{-4}$}\\
            \rule[-1ex]{0pt}{3.5ex}    {$\nu_\mathrm{L,1}$, THz}
                                     & {$1.51 \pm 0.01$}
                                     & {$2.044 \pm 0.003 $}\\
            \rule[-1ex]{0pt}{3.5ex}    {$\gamma_\mathrm{L,1}$, THz}
                                     & {$0.12 \pm 0.01$}
                                     & {$0.23 \pm 0.03$}\\
            \rule[-1ex]{0pt}{3.5ex}    {$\Delta\varepsilon_\mathrm{2}$}
                                     & {$(35.0\pm 2.9) \times 10^{-4}$}
                                     & {$(210 \pm 3) \times 10^{-4}$}\\
            \rule[-1ex]{0pt}{3.5ex}    {$\nu_\mathrm{L,2}$, THz}
                                     & {$2.59 \pm 0.01$}
                                     & {$3.503 \pm 0.002$}\\
            \rule[-1ex]{0pt}{3.5ex}    {$\gamma_\mathrm{L,2}$, THz}
                                     & {$0.42 \pm 0.02$}
                                     & {$0.18 \pm 0.01 $}\\
            \rule[-1ex]{0pt}{3.5ex}    {$\Delta\varepsilon_\mathrm{3}$}
                                     & {$(7.2 \pm 1.0) \times 10^{-4}$}
                                     & {$(13 \pm 2) \times 10^{-4}$}\\
            \rule[-1ex]{0pt}{3.5ex}    {$\nu_\mathrm{L,3}$, THz}
                                     & {$8.11 \pm 0.01$}
                                     & {$7.165 \pm 0.001$}\\
            \rule[-1ex]{0pt}{3.5ex}    {$\gamma_\mathrm{L,3}$, THz}
                                     & {$1.82 \pm 0.01$}
                                     & {$1.15 \pm 0.01$}\\
            \rule[-1ex]{0pt}{3.5ex}    {$\Delta\varepsilon_\mathrm{4}$}
                                     & {--}
                                     & {$(6 \pm 2) \times 10^{-4}$}\\
            \rule[-1ex]{0pt}{3.5ex}    {$\nu_\mathrm{L,4}$, THz}
                                     & {--}
                                     & {$8.88 \pm 0.04$}\\
            \rule[-1ex]{0pt}{3.5ex}    {$\gamma_\mathrm{L,4}$, THz}
                                     & {--}
                                     & {$1.11 \pm 0.01 $}\\
        \end{tabular}
    \label{TAB}
\end{table}

Figures~\ref{FIG:COResponse}
and~\ref{FIG:CO2Response} show the presence of
several pronounced spectral resonances
in the THz--IR range,
whose absorption magnitude
is much higher than the detection limit
evaluated for thicker ices.
For the CO ice, we notice
three Lorentz-like absorption peaks ($\gamma_\mathrm{L}$), and four absorption peaks for the CO$_2$ ice. All the parameters derived from the dielectric permittivity model are summarized in Table\ref{TAB}.

The lower-frequency peaks of CO ($1.51$ and $2.59$~THz) and CO$_{2}$ ($2.04$ and $3.50$~THz) ices are well known from the literature
\citep{JCP.45.11.4359.1966, JCP.46.10.3991.1967, PCCP.16.8.3442.2014, ARAA.52.1.541.2015, AA.629.A112.2019},
while the weaker peaks at higher frequencies ($8.11$~THz for CO; $7.16$ and $8.88$~THz for CO$_{2}$) have not been reported so far, to the best of our knowledge.

The intense low-frequency peaks
($1.51$ and $2.59$~THz of CO ice;
$2.04$ and $3.50$~THz of CO$_{2}$ ice)
are attributed to the intermolecular vibrational modes of
CO and CO$_{2}$ lattices,
somewhat broadened in disordered ices.
On the other hand,
the peaks seen at higher frequencies
($8.11$~THz of CO ice;
$7.16$ and $8.88$~THz of CO$_{2}$ ice)
are substantially broader.
They  may originate from amorphous or polycrystalline ice structure,
representing the so-called Bóson peaks
\citep{PRB.43.6.5039.1991, PRE.61.1.587.2000, CP.41.1.15.2000, PRB.67.9.094203.2003}.
Indeed, an increase in
the anharmonic contribution
(in excess to the potential energy of a crystal)
leads not only to the broadening of
the vibrational resonances,
but also to an increase in dielectric losses
in a wide frequency range.
The latter includes the formation of
additional broad spectral features,
commonly observed in fully or partially disordered media
and referred to as the Boson peaks
\citep{PRL.91.20.207601.2003, RMP.72.3.873.2000, RMP.46.3.465.1974, PR.135.2A.A413.1964}.

\begin{figure}[!t]
    \centering
    \includegraphics[width=1.0\columnwidth]{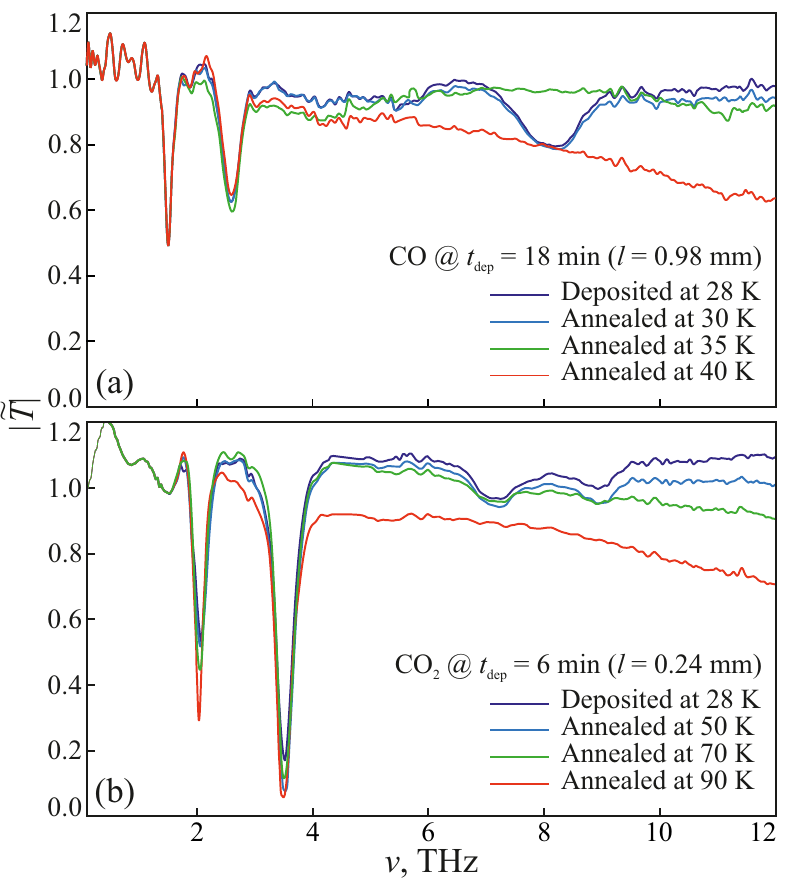}
    \caption{Amplitude of the complex transmission coefficient
    $\left| T \left( \nu \right) \right|$ of
    grown and annealed ices.
    (a)~CO ice deposited at $28$~K
    during $t_\mathrm{dep} = 18$~min ($l = 0.98$~mm),
    and annealed at the temperatures of $30$, $35$, $40$~K.
    (b)~CO$_2$ ice deposited
    at the baseline temperature of $28$~K
    during $t_\mathrm{dep} = 6$~min ($l = 0.24$~mm),
    and annealed at the temperatures of $50$, $70$, $90$~K.}
    \label{FIG:Annealing}
\end{figure}

To verify the discussed nature of the broader high-frequency peaks, we have performed additional experiments with annealing of CO and CO$_{2}$ ices,
aiming to increase the ice order.
For this purpose,
CO ice was deposited for $18$~min
at the baseline temperature of $11$~K
and then annealed at the temperatures of
$30$, $35$, and $40$~K for 15~min;
CO$_{2}$ ice was deposited for $6$~min
at the baseline temperature of $11$~K
and then annealed at $50$, $70$, and $90$~K for 15~min.
The observed FTIR transmission spectra
$\left| \widetilde{T} \left( \nu \right) \right|$
of these ices are plotted
in Fig.~\ref{FIG:Annealing}.
A clear indication that the ice order increases
with the annealing temperature is
a visible deepening
of the vibrational absorption features of
CO and CO$_2$ ices at lower frequencies.
We see that
the high-frequency spectral feature of the CO ice,
which remains practically unchanged at $30$~K,
completely disappears
from the transmission spectrum at $35$~K, where the ice is expected to be nearly fully crystalline \citep{He2021}.
This strongly supports our hypothesis that the high-frequency feature reflects amorphous or polycrystalline structure of the ice at the lower temperatures.
At the same time,
we point out that
increasing the temperature further to $40$~K
leads to a substantial overall reduction of
the transmission at higher frequencies.
For the CO$_2$ ice,
the overall reduction is already seen
at the lowest annealing temperature of $50$~K \citep[where the ice is expected to become polycrystalline, see][]{Mifsud2022, Kouchi2021, He2018}
and, therefore, possible complete disappearance of
the two broad peaks between $70$~K and $90$~K
is obscured.
This overall reduction in
$\left| \widetilde{T} \left( \nu \right) \right|$
is observed to occur at higher temperatures,
and is more pronounced at higher frequencies. Therefore, a likely reason behind  this phenomenon
is an increasing contribution of
the light scattering,
which is induced by growing surface roughness
as the annealing temperature approaches
the desorption temperature of ice
\citep[see][]{AA.628.A63.2019}.

Another important characteristics of
the obtained results
is the actual magnitude of absorption
between the resonance peaks.
From Figs.~\ref{FIG:COResponse}~(b) and~(c)
we see that the measured absorption coefficient of
CO ice between the vibrational peak at $2.59$~THz
and the Boson peak at $8.11$~THz
(as well as to the right from the Boson peak)
is much higher than the value predicted
by the simple model of dielectric permittivity
(red dashed curves).
Since the measured $\alpha$
exceeds the minimal detectable absorption
$\delta \alpha$
(derived from the FTIR data for thicker ices,
see Eq.~\eqref{EQ:DetectableAbsorption}),
such an enhanced absorption
outside the resonance peaks
may have a physical origin.
It can be generally attributed to changes
in the structure and geometry of ice samples,
leading to the formation of broad Boson peaks
and collective excitations
between the spectral resonances,
or inducing the light scattering on ice pores.
On the other hand, we cannot exclude
that for low-absorbing analytes (such as CO ice),
this excess in $\alpha$
may also be due to possible systematic errors of
the FTIR measurements
and imperfections of
the developed data processing methods.

Thus, while it is difficult to quantify
the actual magnitude of $\alpha(\nu)$
between the resonance peaks for CO ices,
we can be confident that it falls
between the measured values
(represented by the green shaded line)
and the model curve (dashed red line).
It is noteworthy that for CO$_2$ ices
(see Fig.~\ref{FIG:CO2Response}),
characterized by
an order-of-magnitude higher absorption of
the resonance peaks,
we observe an excellent agreement of
the modeled and measured $\alpha$
between the resonance peaks.

Undoubtedly, further dedicated analysis of
the CO absorption
between the resonance peaks is needed.
To adequately quantify small values of $\alpha$
in this case,
eliminate any non-physical distortions
from the reference and sample FTIR data,
and properly analyze
the underlying dipole excitations of ices,
one should either focus on studies of
substantially thicker ice sample
\citep{JCP.78.11.6399.1983},
or improve the experimental sensitivity of
the FTIR system. We postpone this work for future studies. 

\subsection{Opacities of dust grains with CO and CO$_{2}$ ice mantles}

\begin{figure}[!t]
   \centering
   \includegraphics[width=1.0\columnwidth]{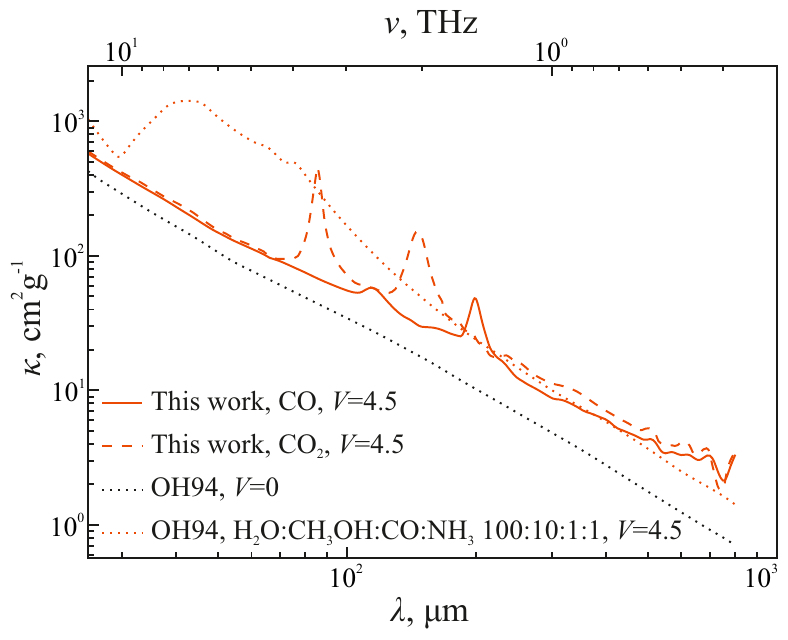}
   \caption{Calculated and reference opacities of astrophysical dust, 
   plotted as a function of the wavelength. Dotted lines labeled with OH94 refer to bare grains as well as to grains with icy mantles by \cite{1994AA.291.943O}. Opacities for grains with pure CO and CO$_2$ icy mantles, computed for optical constants of the present work, are depicted by the solid and dashed lines, respectively. The label "$V$" denotes the volume ratio of the icy mantles to the refractory material, 
   indicating bare grains ($V = 0$, black) and grains with thick mantles ($V = 4.5$, red).
   } 
   \label{FIG:kappa_CO}
    \end{figure}

In Figure~\ref{FIG:kappa_CO} we present the opacity $\kappa$ which is computed for the optical constants of CO and CO$_2$ ices. For direct comparison with available opacity data, results from \cite{1994AA.291.943O} are also included. The opacity is derived by following the procedure described in Sect.~4 and Appendix~C of \citet{AA.629.A112.2019}, the data for the optical constants from our work and the code to calculate the opacity can be found in the online repository\footnote{\url{https://bitbucket.org/tgrassi/compute_qabs}, the project version for this paper is commit:~\texttt{36317b2}. Figure~\ref{FIG:kappa_CO} can be produced by running \texttt{test\_06.py}. Data can be found in \texttt{data/eps\_CO.dat} and \texttt{data/eps\_CO2.dat} files.}. The dotted lines in Fig.~\ref{FIG:kappa_CO} refer to the values of \cite{1994AA.291.943O} for bare grains (black) and for grains covered with thick icy mantles which are composed of water and contain small fractions of other volatile species, H$_{2}$O:CH$_{3}$OH:CO:NH$_{3} = 100$:$10$:$1$:$1$ (red). The labels $V = 0$ and $4.5$ indicate the volume ratio of the ice mantles to the refractory material of grains.\footnote{The volume ratio $V$ is related to the ice thickness $\Delta a$ and the grain radius $a$ via $V=(1+\Delta a/a)^3-1$.} The opacity of grains with thick mantles of pure CO and CO$_{2}$ ices are plotted with the solid and dashed lines, respectively.

A broad absorption feature seen between $\simeq30$ and $\simeq200$~$\mu$m in the opacity curve by \cite{1994AA.291.943O} (dotted line) represents lattice vibrations of the water ice, and therefore is not present in the pure CO or CO$_2$ ice data. The CO opacity curve (solid line) shows two strong features near $1.5$ and $2.6$~THz ($200$ and $115$~$\mu$m), corresponding to the strong absorption peaks in Fig.~\ref{FIG:COResponse}, while the contribution of a weaker Boson peak at about $8.1$~THz cannot be seen. A very similar behavior is observed for the CO$_{2}$ opacity curve (dashed line), clearly showing the signature of two strong absorption peaks in Fig.~\ref{FIG:CO2Response}. At the wavelengths above $200$~$\mu$m, none of the curves show absorption features, and we see a good agreement between different models. Characteristic opacity values at selected wavelengths are given in Table~\ref{TAB2}.

\begin{table}[!b]
    \caption{Opacity $\kappa$ (in units of cm$^2$~g$^{-1}$) calculated at selected wavelengths for CO$_{2}$ and CO ices with a volume ratio of $V = 4.5$.} 
    
    \centering
            \begin{tabular}{ >{\centering\arraybackslash}m{2.0cm}
                        |>{\centering\arraybackslash}m{2.5cm}
                        |>{\centering\arraybackslash}m{2.5cm}}
    \rule[-1ex]{0pt}{3.5ex}    $\lambda$, $\mu$m & CO$_2$ $\kappa$$_{V=4.5}$ & CO $\kappa$$_{V=4.5}$ \\
    \hline  
    \hline

    \rule[-1ex]{0pt}{3.5ex}    $30$     & $417.0$  & $400.4$        \\
    \rule[-1ex]{0pt}{3.5ex}    $50$     & $153.3$  & $149.0$        \\
    \rule[-1ex]{0pt}{3.5ex}    $75$     & $92.2$   & $82.0$         \\
    \rule[-1ex]{0pt}{3.5ex}    $100$    & $71.6$   & $54.6$         \\
    \rule[-1ex]{0pt}{3.5ex}    $115$    & $57.2$   & $57.7$         \\
    \rule[-1ex]{0pt}{3.5ex}    $200$    & $23.9$   & $47.0$         \\
    \rule[-1ex]{0pt}{3.5ex}    $250$    & $16.2$   & $12.2$         \\
    \rule[-1ex]{0pt}{3.5ex}    $350$    & $9.2$    & $7.2$          \\
    \rule[-1ex]{0pt}{3.5ex}    $500$    & $5.3$    & $4.3$          \\ 
    \end{tabular}
    \label{TAB2}
\end{table}

A volume ratio of $V = 4.5$ is a reasonable assumption for thick astrophysical ices composed of several components, such as the ice mixture analyzed by \cite{1994AA.291.943O}. In order to facilitate a comparison, we keep this value also for the opacity curves representing pure CO and CO$_{2}$ ices, but the detectability of their bands will vary according to the actual abundance of the molecules.\footnote{We emphasize that the volume ratio is a free parameter in our opacity code, that can be easily changed in accordance with the context of a problem.} 
The opacity model of pure ices is particularly relevant for regions of the interstellar medium where CO and CO$_2$ molecules are expected to be concentrated in the outer layers of the icy mantle, e.g., outside the respective snow lines in protoplanetary disks, in protoplanetary disks mid-planes or in the center of prestellar cores. 

\section{Conclusion}
\label{SEC:Concl}

In the presented work, we developed and implemented experimentally a new method for quantitative characterization of complex dielectric permittivity of astrophysical ice analogs in a broad THz-IR spectral range. By performing joint processing of TDS and FTIR spectroscopic data, we derived optical constants of CO and CO$_{2}$ ice layers in an extended frequency range of $0.3$--$12.0$~THz, and analyzed the results theoretically using a multiple Lorentzian model.

The extended spectroscopic data of CO and CO$_{2}$ ices show the presence of broad absorption features near $8$~THz, which have not been characterized before to the best of our knowledge. By studying the dielectric response of ices after annealing at different temperatures, we concluded that such bands can be attributed to Boson peaks -- the prominent signatures of disorder in amorphous or polycrystalline media.

Based on the measured broadband THz--IR dielectric response of CO and CO$_{2}$ ices, we estimated and analyzed the opacity of astrophysical dust covered with thick icy mantles of the same molecular composition. These
measurements are necessary to provide a better interpretation of dust continuum observations in star- and planet-forming regions, where catastrophic CO freeze out occurs (in pre-stellar cores and in protoplanetary disk mid-planes) and at the CO and CO$_{2}$ snow lines of protoplanetary disks. 

\begin{acknowledgements}
We gratefully acknowledge the support of the Max Planck Society. This project has received funding from the European Union’s Horizon~2020 research and innovation program under the Marie Skłodowska-Curie grant agreement \#~$811312$ for the Project "Astro-Chemical Origins" (ACO). TDS and FTIR data processing by A.A.G. was supported by the RSF~Project~\#~$22$--$72$--$00092$. The authors thank the referee, S. Ioppolo, for providing insightful and constructive comments and suggestions. 
\end{acknowledgements}

\bibliographystyle{aa} 
\bibliography{REFs} 

\end{document}